\documentstyle[12pt,epsf]{article}
\renewcommand{\bar}[1]{\overline{#1}}

\begin{document}

\begin{flushright}
\hfill{ USM-TH-109}
\end{flushright}

\centerline{\Large \bf Up and Down Quark Contributions to Spin Content }
\centerline{\Large \bf of $\Lambda$ from  Fragmentation}

\vspace{22pt}
\centerline{\bf
Jian-Jun Yang\footnote{e-mail: jjyang@fis.utfsm.cl}$^{a,b}$}

\vspace{8pt}

{\centerline {$^{a}$Department of Physics, Nanjing Normal
University,}}

{\centerline {Nanjing 210097, China}}

{\centerline {$^{b}$Departamento de F\'\i sica, Universidad
T\'ecnica Federico Santa Mar\'\i a,}}

{\centerline {Casilla 110-V, 
Valpara\'\i so, Chile\footnote{Mailing address}}

\vspace{10pt}
\begin{center} {\large \bf Abstract}

\end{center}
We check the $u$ and $d$ quark contributions to the spin content
of the $\Lambda$ by means of  the $q\to\Lambda$ fragmentation and
find that the  $u$ and $d$ quarks of the $\Lambda$ are likely
positively polarized.
 The parton  distributions in the $\Lambda$ are given
 by a successful statistical model which can  reproduce and
 correlate a vast body of polarized and unpolarized
 structure function and parton distribution data of the nucleon.
  With the Gribov-Lipatov relation between the
  quark  distributions and fragmentation functions,
   the longitudinal spin transfer for the $\Lambda$  production in
the polarized charged lepton deep inelastic scattering (DIS)
process and the $\Lambda$-polarization in  the neutrino
(antineutrino) DIS process are predicted. The available
experimental data suggests that{\it{ the  $u$ and $d$ quark
contributions to the spin of the $\Lambda$ are  positive.}} In
addition,  our results provide a collateral evidence for the SU(3)
symmetry breaking in hyperon semileptonic decays of the octet
baryons, which is very important for a deeper understanding of the
proton 'spin crisis'.

\vfill
\centerline{PACS number(s): 14.20.Jn, 13.87.Fh, 13.60.Hb}

\newpage
\section{Introduction}

 In the naive quark model, the $\Lambda$ spin is totally provided by the
strange ($s$) quark, and the $u$ and $d$ quarks are unpolarized.
Based on novel results concerning the proton
spin structure from DIS experiments, it was found
that the $u$ and $d$ quarks of the $\Lambda$ should be
negatively polarized~\cite{Bur93}. This analysis assumes,
however, the SU(3) flavor symmetry for the weak decays in the
baryon octet. Recently, it has been
noticed that the effect of SU(3) symmetry breaking in
 hyperon semileptonic decay (HSD)
 should be  significant~\cite{KimPRD, Manohar98}.
 The effect has been estimated  by the chiral quark
 soliton model~\cite{KimPRD} and the large
 $N_c$ QCD~\cite{Manohar98}. The
 consistent results  were obtained  separately by different approaches.
 The effect of SU(3) symmetry breaking in HSD on the spin content of
 the $\Lambda$ was considered in the chiral quark soliton
 model~\cite{KimAPPB}. It was
 found that the integrated polarized quark
 densities for the $\Lambda$ hyperon should be
 $\Delta U=\Delta D= -0.03 \pm 0.14$ and $\Delta S=0.74  \pm 0.17$
  in the chiral limit case.
 When the strange quark mass correction is added,
 $\Delta U=\Delta D= -0.02 \pm 0.17$ and $\Delta S=1.21  \pm 0.54$.
Therefore, the SU(3) symmetry breaking allows the polarization of
the $u$ and $d$ quarks in the $\Lambda$ to be positive. However,
there is a lack of experimental evidences to check the effect of
the SU(3) symmetry breaking. It should be significant to clarify
the polarization of $u$ and $d$ quarks in the $\Lambda$. First, it
reflects the spin structure of the $\Lambda$ itself. Second,
knowing how the SU(3) symmetry breaking in HSD to affect the spin
content of the $\Lambda$ can help us to have a deeper
understanding of the proton spin 'crisis'. Third, if the $u$ and
$d$ quark contributions to the spin content of the $\Lambda$ are
transferred  into the spin structure of quark fragmentation
functions, they dominate the spin transfer to the $\Lambda$ in the
polarized charged lepton DIS process due to the charge factor for
the $u$ quark. Thus, this subject is very  important for enriching
the knowledge of hadron structure and hadronization mechanism.

Based on a perturbative QCD (pQCD) counting
rules analysis~\cite{countingr,Bro95} and an SU(6)
quark diquark spectator model~\cite{Ma96}, it was
found~\cite{MSSY} that, although the $u$ and $d$
quarks of the $\Lambda$ might be unpolarized or negatively
polarized in the integrated Bjorken range $0 \leq x \leq 1$,
they should be positively polarized at large $x$. However,
neither the pQCD analysis nor the  SU(6) quark diqaurk spectator model can
allow to form a clear judgement whether the $u$ and $d$
quark contributions to the spin content of the $\Lambda$
are zero, negative or positive. Recently, a statistical model
for polarized and unpolarized parton distributions of the nucleon
was presented by Bhalerao {\it{ et
al.}}~\cite{Bhalerao00}. The model can reproduce the almost all
data on the nucleon structure functions $F_2^p(x,Q^2)$,
$F_2^p(x)-F_2^n(x)$ and parton sum rules, which motives us to
extend its application from the nucleon to the $\Lambda$.

In this Letter, we investigate  the $u$ and $d$ quark
contributions to the spin of the $\Lambda$ and check the effect of
the SU(3) symmetry breaking in HSD by means of the statistical
model.
 According to some constraints,
we determine the parton  density  functions (PDFs) for the
$\Lambda$ at an initial scale. Then we relate PDFs to
fragmentation functions at the initial scale by using  the
Gribov-Lipatov  relation between the quark distributions and
fragmentation functions. Finally, we employ the evolved
fragmentation functions to predict the longitudinal spin transfer
to the $\Lambda$ in the polarized charged lepton DIS process and
the $\Lambda$ polarization in the neutrino (antineutrino) DIS
process. With the available experimental data, it is found that
the $u$ and $d$ quark contributions to the spin of the $\Lambda$
are positive, which provides a collateral evidence for the SU(3)
flavor symmetry breaking in HSD.

\section{Quark distributions in the $\Lambda$}

 Recently, it has been found that the
input-scale parton densities in the nucleon  may  be
quasi-statistical in nature~\cite{Flambaum98,Bickerstaff90,Bhalerao96}. With
a statistical model, a vast body of polarized and unpolarized
nucleon structure functions and parton sum rules can be well
described~\cite{Bhalerao00}.  This motivates
 us to apply the same mechanism to other octet hyperons, especially to
 the $\Lambda$.

 Following Ref.~\cite{Bhalerao00}, the parton number density
 $d n^{IMF} /d x$ in
 the infinite-momentum frame (IMF) can be related to the
 density $d n /d E$ in the $\Lambda$ rest frame by

 \begin{equation}
 \frac{dn^{IMF}}{d x} = \frac {M_\Lambda^2 x} {2}
 \int\limits_{x M_\Lambda/2}^
 {M_\Lambda/2} \frac{d E}{E^2} \frac{d n }{dE},\label{dnx}
 \end{equation}
 where $M_\Lambda$ is the mass of the $\Lambda$ and $E$ is
 the parton energy in the $\Lambda$ rest frame.
 It should be pointed out  that Eq.~(\ref{dnx}) is  an assumption even for
massless quarks since it assumes that quarks can be boosted using
a purely kinematic transformation, which is in general not true in
an interacting theory, especially not in a strongly interacting
theory such as QCD. However, the reasonableness  of the model has
been tested by its  successful application  to the prediction of
quark distributions of the nucleon. Extending the model from the
nucleon to the $\Lambda$ can provide an independent check  of the
same mechanism that produces the flavor and spin structure of the
nucleon since the quark structure of the $\Lambda$ is a new
frontier with rich physics.

 In consideration of the effects of the
 finite size of the $\Lambda$, $dn/d E$ can be expressed as
 the sum of the volume, surface and curvature terms,
 \begin{equation}
 dn/dE= g f (E) ( V E ^2 /2 \pi^2+ a R^2 E + b R ),\label{dne}
 \end{equation}
 with the usual Fermi or Bose distribution function
 $f(E)=1/[{\rm{e}}^{(E-\mu)/T} \pm 1 ]$. In (\ref{dne}), $g$ is the spin-color
 degeneracy factor, $V$ is the $\Lambda$ volume and $R$ is the
 radius of a sphere with volume $V$. The parameters $a$ and $b$ in
 (\ref{dne}) have been determined by fitting the structure function data
 for the proton. We choose the same values of them for the $\Lambda$,
 {\it{i.e.}} $a=-0.376$ and $b=0.504$.
 Then, $n_{q(\bar{q})}^{\uparrow(\downarrow)}$ which denotes the
 number of quarks(antiquarks) and spin parallel
 (anti-parallel) to the $\Lambda$ spin can be written as

 \begin{equation}
 n_{q(\bar{q})}^{\uparrow(\downarrow)}=g \int\limits_0^{M_\Lambda/2}
 \frac{V E^2/ 2\pi ^2 + a R ^2 E +b R}
 { {\rm{e}}^{(E-\mu_{q(\bar{q})}^{\uparrow(\downarrow)})/T} + 1} dE
 \end{equation}
 Similarly, the momentum fraction carried by the quark $q$ (antiquark $\bar{q}$) and gluon $G$ can
 be expressed as

 \begin{equation}
 M_{q(\bar{q})}^{\uparrow(\downarrow)}=\frac{4g}{3M_\Lambda}
 \int\limits_0^{M_\Lambda/2}
 \frac{E(V E^2/ 2\pi ^2 + a R ^2 E +b R)}
 { {\rm{e}}^{(E-\mu_{q(\bar{q})}^{\uparrow(\downarrow)})/T} + 1} dE
 \end{equation}

 \begin{equation}
 M_{G}^{\uparrow(\downarrow)}=\frac{4g}{3 M_\Lambda}
 \int\limits_0^{M_\Lambda/2}
 \frac{E(V E^2/ 2\pi ^2 + a R ^2 E +b R)}
 { {\rm{e}}^{(E-\mu_{G}^{\uparrow(\downarrow)})/T} - 1} dE
 \end{equation}
 Due to isospin symmetry, the $u$ and $d$ quarks in
 the $\Lambda$ are expected to be equal.
 Hence, the quark numbers and the parton momentum fractions
 have to satisfy the following five constraints:

 \begin{equation}
 n_u^\uparrow + n_u^\downarrow-n_{\bar{u}}^\uparrow - n_{\bar{u}}^\downarrow
=1, \label{con1}
 \end{equation}

\begin{equation}
 n_u^\uparrow - n_u^\downarrow + n_{\bar{u}}^\uparrow - n_{\bar{u}}^\downarrow
 =\Delta U, \label{con2}
 \end{equation}

 \begin{equation}
 n_s^\uparrow + n_s^\downarrow - n_{\bar{s}}^\uparrow -
  n_{\bar{s}}^\downarrow=1, \label{con3}
 \end{equation}

\begin{equation}
 n_s^\uparrow - n_s^\downarrow + n_{\bar{s}}^\uparrow
  - n_{\bar{s}}^\downarrow
 =\Delta S, \label{con4}
 \end{equation}

\begin{equation}
\sum \limits_{q} (M_q^\uparrow + M_q^\downarrow + M_{\bar{q}}^\uparrow
+ M_{\bar{q}}^\downarrow) + (M_G^\uparrow + M_G^\downarrow)=1, \label{con5}
\end{equation}
where the integrated polarized quark densities $\Delta U$ and
$\Delta S$ are very important inputs for the spin structure of the
$\Lambda$. In order to describe the spin content of the $\Lambda$,
it is necessary to distinguish between
$\mu_{q(\bar{q})}^{\uparrow}$ and $\mu_{q(\bar{q})}^{\downarrow}$.
We assume that the gluon is not polarized at the initial scale and
hence $\mu_{G}^{\uparrow}=\mu_{G}^{\downarrow}=0$. Thus at input
scale, $\Delta G(x) =0$ and the  gluon polarization comes from the
QCD evolution. In addition, it has been noticed that
$\mu_{\bar{q}}^{\uparrow} =-\mu_q^{\downarrow}$ and
$\mu_{\bar{q}}^{\downarrow} =-\mu_q^{\uparrow}$. Therefore, by
solving 5 coupled nonlinear equations (\ref{con1})-(\ref{con5}),
we can determine 5 unknowns, namely $\mu_u^{\uparrow}$,
$\mu_u^{\downarrow}$, $\mu_s^{\uparrow}$, $\mu_s^{\downarrow}$,
and $T$.

Recent analyses show that the  SU(3) flavor  symmetry breaking in
HSD has a significant effect on the  extraction of the
contributions $\Delta U$, $\Delta D$ and $\Delta S$ to the spin of
the octet baryons. In order to check the effect  of the SU(3)
symmetry breaking, we adopt two sets of typical $\Delta U$ and
$\Delta S$ for the $\Lambda$. Set I: $\Delta U=\Delta D= 0.10$ and
$\Delta S=0.74$; Set II: $\Delta U=\Delta D= -0.17$ and $\Delta
S=0.62$. The set-I is built under the guidance of the results
given by the chiral quark model~\cite{KimAPPB} with the SU(3)
symmetry breaking in HSD. The set-II is based on an assumption of
the SU(3) flavor symmetry for the weak decays in the baryon
octet~\cite{Bur93,Bor98}. The corresponding solutions of
$\mu_u^{\uparrow}$, $\mu_u^{\downarrow}$, $\mu_s^{\uparrow}$,
$\mu_s^{\downarrow}$, and $T$ for the two sets of $\Delta U$ and
$\Delta S$ are listed in Table 1.  With these values,  unpolarized
and polarized parton distributions  in the $\Lambda$ can be
obtained directly from (\ref{dnx}).

\vspace{1cm}

\centerline{Table 1~~Chemical potentials ($\mu$) and temperature
($T$) (in MeV).}

\vspace{0.3cm}

\begin{footnotesize}
\begin{center}
\begin{tabular}{|c||c|c|c||c|c|c|c|c|}\hline
 Set & $\Delta U$ & $\Delta D$ & $\Delta S$
 &$\mu_u^{\uparrow}$ &$\mu_u^{\downarrow}$
 &$\mu_s^{\uparrow}$ &$\mu_s^{\downarrow}$ & $T$\\ \hline
 I & 0.10 & 0.10 & 0.74 & 80.5 & 65.8 & 127.3 & 19.0 & 73.8\\
\hline II & -0.17 & -0.17 & 0.62 & 60.7 & 85.6 &
118.5 & 27.7 & 74.0\\ \hline
\end{tabular}
\end{center}
\end{footnotesize}
\vspace{0.5cm}

\section{ Fragmentation functions for the $\Lambda$ polarization}

Unfortunately, we can not check the obtained parton distributions
of the $\Lambda$  by means of structure functions in DIS
scattering since the $\Lambda$ can not be used as a target due to
its short life time. Also one obviously can not produce a beam of
charge-neutral $\Lambda$. What one can check with  experiments is
the  quark to $\Lambda$ fragmentation, and therefore one needs a
relation between  the quark distributions and fragmentation
functions. Recently, there has been progress in understanding the
quark to $\Lambda$ fragmentation~\cite{MSSY} by using the
Gribov-Lipatov (GL) relation~\cite{GLR}
\begin{equation}
D_q^h(z) \sim z \, q_h(z)~ \label{GLR}
\end{equation}
in order to connect the fragmentation functions with the
distribution functions. This relation, where $D_q^h(z)$ is the
fragmentation function for a quark $q$ splitting into a hadron $h$
with longitudinal momentum fraction $z$, and $q_h(z)$ is the quark
distribution of finding the quark $q$ inside the hadron $h$
carrying a momentum fraction $x=z$, is only known to be valid near
$z \to 1$ at an energy scale $Q^2_0$ in leading order
approximation~\cite{BRV00}. However, with the GL relation,
predictions of $\Lambda$ polarizations~\cite{MSSY} based on quark
distributions of the $\Lambda$ in the SU(6) quark diquark
spectator  model and in the pQCD based counting rules analysis,
have been found to be supported by all available data from
longitudinally polarized $\Lambda$ fragmentation in
$e^+e^-$-annihilation~\cite{ALEPH96,DELPHI95,OPAL97}, polarized
charged lepton DIS~\cite{HERMES,E665}, and most recently, neutrino
(antineutrino) DIS~\cite{NOMAD}. Thus we still use (\ref{GLR}) as
an Ansatz to relate the quark fragmentation functions for the
$\Lambda$ to the corresponding quark distributions   at an initial
scale. Then, the quark fragmentation functions are evolved from
the initial scale to the experimental energy scale. We used the
evolution package of Ref.~\cite{Miyama94} suitable modified for
the evolution of fragmentation functions in leading order, taking
the initial
 scale $Q_0^2=M_\Lambda^2$ and $\Lambda_{QCD}=0.3~\rm{GeV}$.
In Fig.~\ref{a02f1}, the set-I (solid lines) and set-II (dashed lines)
 fragmentation functions are presented at two different scales.
The fragmentation functions at the initial scale, which are
related to the corresponding quark distributions via the GL
relation, are presented with thin lines. The fragmentation
functions at $Q^2=10~\rm{GeV}^2$ are shown in thick lines. From
Fig.~\ref{a02f1}(b) where the solid and dashed lines almost
overlap, we find that the two sets of unpolarized $u$ quark to
$\Lambda$ fragmentation functions are almost the same, although
the corresponding polarized fragmentation functions are very
different(see Fig.~\ref{a02f1}(d)). The thin and thick lines in
Fig.~\ref{a02f1}(e)-(f) almost overlap, which indicates that the
$Q^2$ dependence in the spin structure of the fragmentation
functions  is very weak.

\begin{figure}
\begin{center}
\leavevmode {\epsfysize=8cm \epsffile{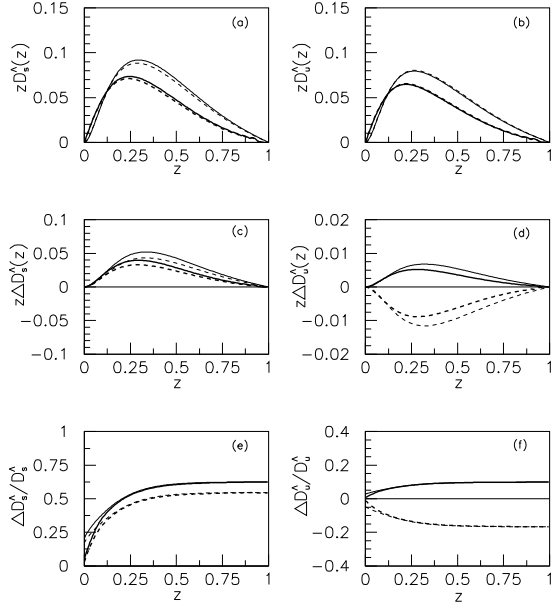}}
\end{center}
\caption[*]{\baselineskip 13pt The quark to $\Lambda$
fragmentation functions. The solid and dashed lines  are
 for set-I and set-II (see Table 1), respectively.
 The thin and thick lines represent the fragmentation functions
 at the initial scale and  at $Q^2=10~\rm{GeV}^2$, respectively.
 Note that the thin and thick lines in (e) and (f) almost overlap.}\label{a02f1}
\end{figure}

We need some experimental data to check the obtained quark
fragmentation functions  for the $\Lambda$. There has been some
recent progress in the measurements of polarized $\Lambda$
production. The longitudinal $\Lambda$ polarization in $e^+e^-$
annihilation at the Z-pole was observed by several
collaborations~\cite{ALEPH96,DELPHI95,OPAL97}. The HERMES
collaboration at DESY and the E665 Collaboration at
FNAL~\cite{E665}  reported their results for the longitudinal spin
transfer to the $\Lambda$ in polarized positron DIS~\cite{HERMES}.
Very recently, the measurement results  of $\Lambda$ polarization
in charged current interactions was obtained in
NOMAD~\cite{NOMAD}. Based on the available experimental data, we
can check the prediction for the $\Lambda$ polarization,
especially in this work, the $u$ and $d$  quark contributions to
the spin content of the $\Lambda$. In $e^+e^-$ annihilation at the
Z-pole, the $\Lambda$ polarization is dominated by strange quark
fragmentation. In order to show the $u$ and $d$  quark
contributions to the spin content of the $\Lambda$, we focus our
attention on the $\Lambda$ electroproduction in which the
longitudinal spin transfer to the $\Lambda$   is dominated from
the $u$ quark due to the charge factor for the $u$ quark.

 For a longitudinally polarized charged
lepton beam and an unpolarized target, the $\Lambda$ polarization
along its own momentum axis is given in the quark
parton model by~\cite{Jaf96}
\begin{equation}
P_{\Lambda}(x,y,z) = P_B D(y)A^{\Lambda}(x,z)~,
\label{PL}
\end{equation}
where $P_B$ is the polarization of the charged lepton beam, which
is of the order of 0.7 or so~\cite{HERMES,E665}. $D(y)$, whose
explicit expression is
\begin{equation}
D(y)=\frac{1-(1-y)^2}{1+(1-y)^2},
\end{equation}
is commonly referred to as the longitudinal depolarization factor
of the virtual photon with respect to the parent lepton, and
\begin{equation}
A^{\Lambda}(x,z)= \frac{\sum\limits_{q} e_q^2 [q^N(x,Q^2) \Delta
D_q^\Lambda(z,Q^2) + ( q \rightarrow \bar q)]}
{\sum\limits_{q} e_q^2 [q^N (x,Q^2)
D^\Lambda_q(z,Q^2) + ( q \rightarrow \bar q)]}~,
\label{DL}
\end{equation}
is the longitudinal spin transfer to the $\Lambda$. Here $y=\nu/E$
is the fraction of the incident lepton's energy that is
transferred to the hadronic system by the virtual photon.
 In Eq.~(\ref{DL}), $q^N(x,Q^2)$,  the quark distribution of the proton,
will be adopted as  the CTEQ5 set 1 parametrization
form~\cite{CTEQ5} in our numerical calculations.

\begin{figure}
\begin{center}
\leavevmode {\epsfysize=3.5cm \epsffile{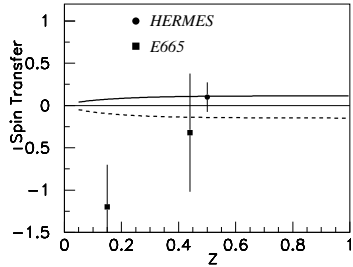}}
\end{center}
\caption[*]{\baselineskip 13pt The $z$-dependence of the $\Lambda$
spin transfer in electron or positron (muon) DIS.
The solid and dashed  curves are
 for set-I and set-II (see Table 1), respectively. Note
that for the HERMES data the $\Lambda$ polarization is measured
along the virtual-photon momentum, whereas for E665 it is measured
along the virtual-photon spin. The averaged value of the Bjorken
variable is chosen as $x=0.1$ (corresponding to the HERMES
averaged value) and the calculated result is not sensitive to a
different choice of $x$ in the small $x$ region (for example,
$x=0.005$ corresponding to the E665 averaged value).
$Q^2=4~\rm{GeV}^2$ is used and the $Q^2$ dependence of the result
is very weak. }\label{a02f2}
\end{figure}

Our predictions with the two sets of fragmentation functions are
shown in Fig.~\ref{a02f2}. We find that both predictions are
compatible in medium $z$ to the data on the longitudinal spin
transfer for the $\Lambda$ production in the polarized charged
lepton DIS process. Although the precision of the data  is not
sufficient to draw  a clear distinction between the two
predictions, it seems that the data favors somewhat the set-I
prediction, {\it{i.e.}}, the $u$ and $d$ quarks are likely
positively polarized in the $\Lambda$. This is consistent with the
analysis by Florian, Stratmann and
 Vogelsang~\cite{Flo98b}. In order to convert the spin
structure of the $\Lambda$ into predictions for future
experiments, authors of Ref.~\cite{Flo98b} made  a QCD analysis of
the polarized $\Lambda$ fragmentation function within three
different scenarios. Scenario 1 corresponds to the SU(6) symmetric
non-relativistic quark model, according to which the $u$ and $d$
quarks of the $\Lambda$ are not polarized; Scenario 2 is based on
an SU(3) flavor symmetry analysis and leads to the prediction that
the $u$ and $d$ quarks of the $\Lambda$ are
 negatively polarized. Scenario 3 is built on the assumption
 that all light quarks contribute equally to the
 $\Lambda$ polarization.
  It is very interesting that the best agreement with data was
 obtained within scenario 3~\cite{HERMES}, which supports
 our present observation  that {\it{ the $u$ and $d$ quarks in
 the $\Lambda$ are likely positively polarized as well as the strange quark}}.

In addition, the scattering of a neutrino beam on a  hadronic
target provides  a source of polarized quarks with specific flavor
structure, and this particular property makes the neutrino
(antineutrino) process an ideal laboratory to study the
flavor-dependence of quark to hadron fragmentation functions,
especially in the polarized case. We find that the $\Lambda$
polarization in the neutrino (anti-neutrino) DIS process can also
be used to check the $u$ and $d$  quark contributions to the spin
content of the $\Lambda$. The longitudinal polarizations of the
$\Lambda$  in its momentum direction, for the  $\Lambda$ in the
current fragmentation region can be expressed as,
\begin{equation}
P_\nu^\Lambda (x,y,z)=-\frac{[d(x)+\varpi s(x)] \Delta D _u^\Lambda (z)
-( 1-y)
^2 \bar{u} (x) [\Delta D _{\bar{d}}^\Lambda (z)+\varpi
\Delta D_{\bar{s}}^\Lambda (z)]}
{[d(x)+\varpi s(x)] D_u ^\Lambda
(z) + (1-y)^2 \bar{u} (x) [D _{\bar{d}}^\Lambda (z)+\varpi
D_{\bar{s}}^\Lambda (z)]}~,\label{neu1}
\end{equation}

\begin{equation}
P_{\bar{\nu}}^\Lambda (x,y,z)=-\frac{( 1-y) ^2 u (x) [\Delta D
_d^\Lambda (z)+\varpi \Delta D
_s^\Lambda (z)]-[\bar{d}(x)+\varpi \bar{s}(x)] \Delta
D _{\bar{u}}^\Lambda (z)}{(1-y)^2
u (x) [D _d^\Lambda (z)+\varpi D
_s^\Lambda (z)]+[\bar{d}(x)+\varpi \bar{s}(x)] D_{\bar{u}} ^\Lambda (z)}~,
\label{neu2}
\end{equation}
where the terms with the factor $\varpi=\sin^2 \theta_c/\cos^2
\theta_c$ ($\theta_c$ is the Cabibbo angle) represent Cabibbo
suppressed contributions.

\begin{figure}
\begin{center}
\leavevmode {\epsfysize=4cm \epsffile{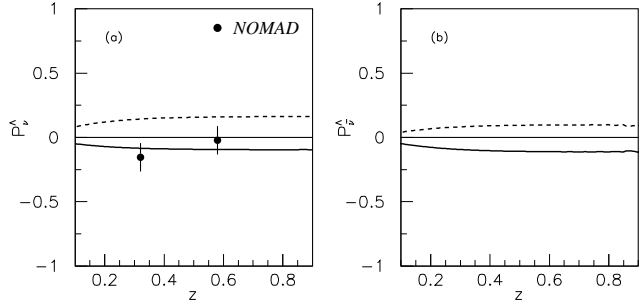}}
\end{center}
\caption[*]{\baselineskip 13pt The prediction of $z$-dependence
for the $\Lambda$  polarization in the neutrino (antineutrino) DIS
process. The solid and dashed curves are for set-I and set-II
(see Table 1), respectively. We adopt the CTEQ5 set 1 quark
distributions~\cite{CTEQ5} for the target proton at
$Q^2=4$~GeV$^2$ with the Bjorken variable $x$ integrated over
$0.02 \to 0.4$ and $y$ integrated over $0 \to 1$. }\label{a02f3}
\end{figure}

The NOMAD data~\cite{NOMAD} on the $\Lambda$ polarization in the
neutrino DIS process, which has much smaller errors than the data
on the  longitudinal spin transfer to the $\Lambda$  in polarized
charged lepton DIS process and shows a weak dependence on $z$, can
help us to distinguish two sets of predictions.
 In Fig.~\ref{a02f3} we present our predictions for the $\Lambda$
 polarization in the neutrino (antineutrino) DIS
process and find that the set-I prediction is  much closer to the
experimental data than the set-II prediction.
The data supports again the set-I that {\it{ the integrated polarized
 $u$ and $d$ quark densities for the $\Lambda$ are positive.}}

\section{Summary and Conclusion}

We constrained  the quark distributions  of the $\Lambda$ at an
initial scale with two sets of typical $\Delta U$ and $\Delta S$
for the $\Lambda$. By means of the statistical model, we
calculated quark distributions for a $\Lambda$ in the rest frame
and then used free boost transformations to relate the rest frame
results to the IMF and made predictions about PDFs. Furthermore,
we  used the GL relation as an Ansatz to relate fragmentation
functions to the corresponding PDFs at the initial scale. Finally,
employing the evolved quark fragmentation functions of the
$\Lambda$, we calculated the longitudinal spin transfer to the
$\Lambda$ in the polarized charged lepton DIS process and the
$\Lambda$ polarization in the neutrino (antineutrino) DIS process.
It is found that the available polarized  charged lepton DIS
experimental data is not sufficient for us to distinguish two sets
of the fragmentation functions, although the data  favors somewhat
the set-I prediction. Fortunately, the very recent NOMAD data on
the $\Lambda$ polarization in the neutrino DIS process, which has
small errors, allow to draw a clear distinction between two sets
of predictions. The experimental data favors obviously the set-I
prediction, which indicates that {\it{the $u$ and $d$ quark
contributions to the spin content of the $\Lambda$ are likely
positive}}.

In addition, our results reflect the importance of the SU(3)
symmetry breaking in HSD. Recently, various theoretical analyses
have arrived at the same conclusion that the flavor SU(3) symmetry
breaking in HSD significantly affects the contributions $\Delta
U$, $\Delta D$ and $\Delta S$ of the light quarks to the spin of
the octet baryons. However, there is a lack of experimental
evidences to support the above conclusion. In our analysis, the
set-I $\Delta U$ and $\Delta S$ for the $\Lambda$ are allowed due
to the SU(3) symmetry breaking in HSD and the set-II $\Delta U$
and $\Delta S$ are based on the assumption of the SU(3) flavor
symmetry for the weak decays in the baryon octet. The set-I
prediction is much better than the set-II prediction in explaining
the NOMAD data on the $\Lambda$ polarization in neutrino DIS
process. Thus, our results provide a collateral evidence  for  the
SU(3) symmetry breaking in HSD, which is very important for a
deeper understanding of the proton 'spin crisis'.

We would like to mention that  our present knowledge on the
$\Lambda$ fragmentation functions is still poor and there are many
unknowns to be explored before we can arrive at some definite
conclusion on the quark spin structure of the $\Lambda$. First,
what one actually measures in experiments are  quark to $\Lambda$
fragmentation functions, and we used the GL relation Eq.~(11) in
order to relate the quark distributions inside $\Lambda$ with the
fragmentation functions of the same flavor quark to $\Lambda$.
 However, such a relation is only known to be valid in large $z$ region,
  and using it down to very small $z$ is questionable.
Second, there are still uncertainties on the quark distributions
of the $\Lambda$ given by the statistical model  since some
assumptions are underlying the model. Despite the experimental
uncertainties, it seems that the experimental data on the
longitudinal spin transfer to the $\Lambda$  in the polarized
charged lepton DIS process shows a  strong dependence on $z$,
especially in low $z$ region. At  the moment, it seems to be
difficult to understand such a rather strong $z$ dependence in low
$z$ region with the available models~\cite{MSSY,Flo98b}, although
the models can provide predictions which are compatible to the
data in medium $z$ range. Thus, we need to improve our prediction
with respect to the $z$ dependence. In order to provide a set of
more realistic fragmentation functions for the $\Lambda$, further
studies, such as revising the GL relation and improving the quark
distributions of the $\Lambda$, are in progress. On  the other
hand, experimentally, the high statistics investigation of
polarized $\Lambda$ production is one of the main future goals of
the HERMES Collaboration which will improve their detector for
this purpose by adding so called Lambda-wheels. The physics of
$\Lambda$ polarization has been regarded as a strongly emphasized
project in the COMPASS experiment~\cite{CSM}. Many efforts, both
theoretically and experimentally, are being made in order to
reduce the uncertainties in the spin  structure of the $\Lambda$
since the subject is crucial important for enriching the knowledge
of hadron structure and hadronization mechanism.

{\bf Acknowledgments: } This work is inspired by my cooperative
work with Bo-Qiang Ma, Ivan Schmidt, and Jacques Soffer. I would
like to express my great thanks to them for the enjoyable
collaboration and encouragement from them. I also thank for
A. Sch\"{a}fer's helpful discussions and  N. G.
Kelkar's correspondence about the gluon in the statistical model. In
addition, this work is partially supported by National Natural
Science Foundation of China under Grant Number 19875024 and by
Fondecyt (Chile) project 3990048.

\end{document}